\documentstyle[aps,prb,preprint]{revtex}

\begin{document}
\draft
\author{P. Garc\'{\i}a-Gonz\'{a}lez and R. W. Godby}
\address{Department of Physics, University of York, Heslington,
York YO10 5DD, United Kingdom.}
\title{Self-Consistent Calculation of Total Energies of the Electron Gas Using
Many-Body Perturbation Theory}
\date{\today }
\maketitle

\begin{abstract}
The performance of many-body perturbation theory for calculating
ground-state properties is investigated. We present fully numerical results
for the electron gas in three and two dimensions in the framework of the $GW$
approximation. The overall agreement with very accurate Monte Carlo data is
excellent, even for those ranges of densities for which the $GW$ approach is
often supposed to be unsuitable. The latter seems to be due to the
fulfilment of general conservation rules. These results open further
prospects for accurate calculations of ground-state properties circumventing
the limitations of standard density functional theory.
\end{abstract}


\section{INTRODUCTION}

Many-body perturbation theory (MBPT), particularly in Hedin's $GW$
approximation \cite{He65}, has been used extensively to calculate
quasiparticle (QP) energies and spectra of a wide variety of electron
systems \cite{Hy85}. The $GW$ method offers a simple way to determine the
one-electron Green's function, $\widehat{G}$, from which the QP properties
can be easily extracted. However, $\widehat{G}$ also contains information
about ground-state properties: the expectation value of any one-particle
operator can be expressed in terms of $\widehat{G}$, and by using the
Galitskii-Migdal formula \cite{GM58} the total energy may also be obtained.
Nonetheless, the capability of a Green's function MBPT to provide reliable
ground-state energies has not been fully explored so far. The few available
results are restricted to the spin-unpolarized homogeneous electron gas
(HEG) in the range of metallic densities (i.e., $r_{{\rm s}}=2\sim 5$ a.u.) 
\cite{Lu67,Sh96,Ho99}. These investigations suggest that the $GW$ approach
could produce accurate electron total energies, but a deeper study is needed
to provide an overall assessment of this issue. This is an important
question because many of the limitations of the usual implementations of
density functional theory (DFT) \cite{KS65}, can be circumvented by using
MBPT total-energy calculations. Indeed, well-recognized failures of the DFT
in its usual approximations (for instance when studying van der Waals forces 
\cite{Do99}, chemical reactions \cite{GM97}, defects in semiconductors \cite
{Le99}, or quasi-two dimensional systems \cite{Ki00}) are mainly due to the
limited account of non-local effects that, on the contrary, are included in
the $GW$ approximation. On the other hand, quantum Monte Carlo (QMC) methods
are being applied to more and more systems \cite{Ne00}, but they require a
large computational effort. In this context, MBPT has to be regarded as a
good candidate to supersede standard DFT schemes [like the local density
(LDA) and the generalized gradient approximations] without implying a
prohibitive computational task. To provide insights into the above points,
in this Paper we present $GW$ results for the ground-state properties of the
three-dimensional HEG (covering a broad range of densities in both
spin-unpolarized and fully spin-polarized phases) and of the two-dimensional
HEG. To do so we have used the {\em space-time} numerical procedure
developed by Rojas {\em et al.} \cite{Ro95}. This method permits efficient
and stable computation of the full self-energy operator $\widehat{\Sigma }$
and the corresponding Green's function $\widehat{G}$ \cite{St99}, with the
precision needed for a converged evaluation of the total energy.

\section{THEORY}

In MBPT, the Green's function and the self-energy of a system of $N$
electrons under a external potential $v_{{\rm ext}}\left( {\bf r}\right) $
are linked through the Dyson equation 
\begin{equation}
\widehat{G}^{-1}\left( \omega \right) =\widehat{G}_{0}^{-1}\left( \omega
\right) -\left[ \widehat{\Sigma }\left( \omega \right) +\left( \Delta
v-\Delta \mu \right) \widehat{1}\right] \;,  \label{1}
\end{equation}
where the usual matrix operations are implied. $\widehat{G}_{0}\left( \omega
\right) $ is the Green's function of a fictitious system of $N$
non-interacting electrons under the potential $v_{0}\left( {\bf r}\right)
+\Delta \mu $, $\Delta v=v_{{\rm H}}+v_{{\rm ext}}-v_{0}$ ($v_{{\rm H}}$
being the exact Hartree potential), and $\Delta \mu $ is a constant that
aligns the chemical potential of the fictitious system with the actual one, $%
\mu $.\cite{Not1}. In the $GW$ framework, $\widehat{\Sigma }$ is
approximated by 
\begin{equation}
\Sigma \left( 1,2\right) =i\,G\left( 1,2^{+}\right) W\left( 1,2\right) \;,
\label{2}
\end{equation}
where the labels 1,2 symbolize space-time coordinates. $\widehat{W}$ is the
screened Coulomb potential which is related to the bare Coulomb interaction $%
w$ and the polarizability $\widehat{P}$ by 
\begin{equation}
\widehat{W}\left( \omega \right) =\widehat{w}+\widehat{w}\,\widehat{P}\left(
\omega \right) \,\widehat{W}\left( \omega \right) \;.  \label{3}
\end{equation}
Finally, under the $GW$ approach we have \cite{Not2} 
\begin{equation}
\widehat{P}\left( 1,2\right) =-2i\,G\left( 1,2\right) \,G\left(
2,1^{+}\right) \;.  \label{4}
\end{equation}
Eq. \ref{1}-\ref{4} may be solved iteratively to self-consistency. We note
that the choice of the fictitious non-interacting system in (\ref{1}) is
arbitrary because the differences arising from different $\widehat{G}_{0}$s
are cancelled out by the terms $\Delta v$ and $\Delta \mu $. Also, by
including the shift $\Delta \mu $ we guarantee that at any step of the
iteration, $\widehat{G}$ verifies several exact properties that have to be
verified by any realistic Green's function \cite{Sh97}, ensuring a smooth
and stable convergence of the iterative process. Furthermore, the Hartree
potential has to be updated after each iteration, and this is done by
calculating the electron density $n\left( {\bf r}\right) $ from $\widehat{G}$%
. To evaluate $\Delta \mu $ we need to obtain the chemical potential that
is, by definition, the energy of the highest occupied QP energy, calculated
at each iteration in terms of the self-energy by solving the QP
Schr\"{o}dinger equation.

The above set of equations defines a {\em conserving approximation} in the
Baym-Kadanoff sense \cite{BK61}. One consequence is that the total number of
particles given by the self-consistent $GW$ Green's function does not change
when an external perturbation acts on the system. Besides, it gives the
right number of particles for the HEG \cite{Ka62}. The correctness of the
number of particles for an arbitrary inhomogeneous system can thus be
inferred by regarding the system as the result of an adiabatic transform of
the HEG. Another characteristic of a conserving approximation is the absence
of ambiguities among different expressions for calculating the total energy.
This suggests that a MBPT evaluation of ground-state properties should
employ a conserving approximation such as self-consistent $GW$.

However, routine $GW$ calculations are mainly concerned with the QP
properties of real materials and do not attempt self-consistency. Indeed,
self-consistency implies a worsening in the description of the QP spectrum
rather than an improvement \cite{Ho99,BH96,SE98}. Hence, the usual non
self-consistent (and non-conserving) $GW$ approach (that we shall denoted as 
$G_{0}W_{0}$, whereas $GW$ will stand for the fully self-consistent solution
of Hedin's equations) is clearly preferred in a spectral context. This
failing of $GW$ can be understood in terms of the spectral properties of $%
\widehat{G}$, but our present interest is very different: issues as those
described in the previous paragraph are by far more important than the
concrete shape of $\widehat{G}$. Of course these two aspects are not
independent, but the development and application of a conserving theory
giving at the same time an accurate description of QP spectra is an unsolved
formidable challenge.

In $G_{0}W_{0}$, the self-energy is approximated by Eq. \ref{2} supposing
that $\widehat{G}=\widehat{G}_{0}$, whereas the screened Coulomb potential
was obtained from (\ref{4}) and (\ref{3}) once (i.e., $\widehat{W}$ has been
calculated in a RPA fashion). Eventually, the Dyson equation is solved
taking into account that $\Delta v\simeq v_{{\rm LDA}}$, where $v_{{\rm LDA}%
} $ is the LDA exchange-correlation (XC) potential. Although the term $%
\Delta \mu $ is often neglected at this stage, in this Paper we will keep
this contribution for the reasons explained above. Partial self-consistency
(denoted as $GW_{0}$) may be achieved by keeping the screened Coulomb
potential $\widehat{W}_{0}$ obtained from a $G_{0}W_{0}$ procedure and,
hence, by solving only (\ref{1}) and (\ref{2}) iteratively \cite{Sh96,BH96}.
Although the $GW_{0}$ is not a conserving approximation, it gives the right
number of particles for the HEG (so meeting an important physical point of
the $GW$ scheme) \cite{Ho99}. On the other hand, its description of QP
properties seems to be only marginally worse than $G_{0}W_{0}$ \cite{BH96}.
Finally, there are many other schemes for implementation of the $GW$
equations, but essentially they are focused on the choice of $\widehat{G}%
_{0} $ for non self-consistent calculations, and so are meaningless in an
homogeneous system.

As mentioned above, we will apply the space-time method \cite{Ro95} to solve
the $GW$ equations. Each one is written in the most favorable spatial
representation (reciprocal or real), going from one to other using Fourier
transforms. However the most important issue is the evaluation of the
dynamical dependence on imaginary time or frequency domain. To calculate
ground-state properties, a contour deformation avoids the need to obtain $%
\widehat{G}$ for real frequencies. Concretely, the expectation value of any
one-particle operator $\widehat{b}$ is given by 
\begin{equation}
\left\langle \widehat{b}\right\rangle =\frac{2}{\pi }{\rm Im}\int_{C}d\omega
\,{\rm Tr}\left[ \widehat{b}\widehat{G}\left( \omega \right) \right] ,
\label{5}
\end{equation}
where the frequency is measured from the chemical potential $\mu $, and $C$
is the integral path in the complex frequency plane equal to the circular
arc $\gamma $ from $\omega =-\infty $ to $\omega =-i\infty $ together with
the negative imaginary axis. In the same notation, the Galitskii-Migdal
formula for the total energy reads 
\begin{equation}
E=\frac{1}{\pi }{\rm Im}\int_{C}d\omega \,{\rm Tr}\left[ \left( \omega +%
\widehat{h}_{0}\right) \widehat{G}\left( \omega \right) \right] ,  \label{6}
\end{equation}
$\widehat{h}_{0}$ being the one-electron hamiltonian with potential $v=v_{%
{\rm ext}}-\mu $. To deal with the evaluation of (\ref{5}) and (\ref{6}) we
write the Green's function as $\widehat{G}=\widehat{G}_{{\rm X}}+\delta 
\widehat{G}$, $\widehat{G}_{{\rm X}}$ being the solution of the Dyson's
equation (\ref{1}), but substituting the full self-energy $\widehat{\Sigma }%
\left( \omega \right) $ by its frequency-independent part $\Sigma _{{\rm X}%
}\left( 1,2\right) =iG\left( 1,2^{+}\right) w\left( 1,2\right) $. Hence, the
frequency integrals can be split up in two parts. The contribution due to $%
\widehat{G}_{{\rm X}}$ is evaluated analytically, whereas for the remainder
the only non-zero contribution arises from the imaginary axis, which is
amenable for numerical calculation. We have used Gauss-Legendre (GL) grids
for imaginary times and frequencies, and the contributions due to points
outside the GL grids are treated in an analytical fashion in accordance with
the overall numerical procedure given in Ref. \cite{St99}. Usually, GL grids
with 128 points suffice for well-converged results (better than 1 mHa). For
the HEG, matrix inversions are not needed, and a fully self-consistent
resolution of the $GW$ equations only takes typically a few seconds on a
standard workstation.

\section{RESULTS AND DISCUSSION}

Using different $GW$ schemes, we have obtained the XC energy per particle, $%
\varepsilon _{{\rm XC}}$ (defined as the difference between the energies of
the interacting and non-interacting systems) for the spin-unpolarized ($%
\zeta =0$) HEG (Table \ref{tab1} and Fig. \ref{fig1}). We first compare our
numerical results with the two of von Barth and Holm \cite{Ho99}. A small
discrepancy ($1\sim 2$ mHa) appears, but it is consistent with the error bar
(about 3\%) of the semi-analytical procedure carried out by these authors 
\cite{Hi97}. Focusing on the accuracy of the MBPT procedure, we can see that
the agreement between the essentially exact diffusion Monte Carlo (QMC) \cite
{CA80,OB99} and the self-consistent $GW$ is almost perfect in the limit of
high densities. This is not a surprise because the exchange is treated
exactly by the $GW$ and it is dominant in this range of densities. However,
the quality of the $GW$ energies is striking for intermediate and low
densities, where the many-body effects not included in the $GW$ framework
might be expected to be evident. Partial self-consistency ($GW_{0}$) yields
slightly inferior results, though the differences are no more than a few
mHa. The worst results (but, in any case, with errors no greater than 10 mHa
for metallic densities) are provided by the non-self-consistent $G_{0}W_{0}$
procedure. $G_{0}W_{0}$ underestimates the total energy, and by achieving
partial self-consistency, the spectral weight in the Green's function is
blue-shifted, so increasing the total energy. After full self-consistency,
such shift is slightly larger, but the kinetic energy is smaller than in $%
GW_{0}$. The presence of these two opposite trends explains the small
differences between $GW$ and $GW_{0}$. We have also included (for
comparison) the corresponding RPA values \cite{Bi82}. Note that the RPA
dielectric function is the same than the $G_{0}W_{0}$ one, and the huge
discrepancies between them arise from the different ways in which the
evaluation of the total energy is performed.

All the above trends also apply for the fully spin-polarized ($\zeta =1$)
HEG (see Fig. \ref{fig2}), but in this case the errors in the $GW$ energies
are marginally greater. (Although not at all the objective of this Paper, it
is interesting to note that using the self-consistent $GW$, the paramagnetic
phase is more stable than the ferromagnetic one up to $r_{{\rm s}}=15\sim 20$%
, in fair agreement with the QMC value \cite{OB99} of $r_{{\rm s}}=25\sim 30$%
, despite the low density of the system).

Our results for the two-dimensional (2D) HEG are also shown. In 2D systems,
correlation effects are much more important; in other words, the diagrams
that are neglected in the $GW$ scheme play a relevant role in these
low-dimensional problems. As a consequence we cannot expect here extremely
accurate results using the $GW$ approximation. However, as we can see in
Fig. \ref{fig3}, the $GW$ gives energies near the QMC values \cite{TC90},
resolving partially the inaccuracy of the $G_{0}W_{0}$ approach and greatly
improving the RPA energies. We note that in the limit of low densities, $%
GW_{0}$ fits the QMC results slightly better than $GW$.

Finally, it is now very well known that $G_{0}W_{0}$ does not give the right
number of particles for an inhomogeneous system \cite{Sh97,RG99}. However
there were certain doubts whether it recovers the right density of the HEG
or not. The use of the space-time method allows us to affirm that $%
G_{0}W_{0} $ does not give exactly the number of particles also in the
homogeneous limit. Whereas the exact density and the $G_{0}W_{0}$ one are
indistinguishable up to $r_{{\rm s}}=4$, for $r_{{\rm s}}=5$ the $G_{0}W_{0}$
overestimates the density by 0.3\%. The deviation increases when going into
the low density region, being 1.7\% for $r_{{\rm s}}=10$, and 6.1\% for $r_{%
{\rm s}}=20$.

In summary, we have studied the performance of the $GW$ approximation for
the evaluation of ground-state properties. The accuracy of the results is
correlated with the fulfillment of conservation rules (that can be achieved
by using a self-consistent $GW$ scheme), and the approximations inherent in
the $GW$ scheme have much less importance than when calculating QP
properties. The dynamical dependences in the $GW$ equations are easy to
handle using a representation in imaginary time and frequency, that may be
straightforwardly generalized to arbitrary inhomogeneous systems. Hence, the
results presented here can be the point of departure for future accurate
evaluations of ground-state properties of electron systems without the
limitations of DFT and the complexity of QMC.

\section*{ACKNOWLEDGMENTS}

We acknowledge illuminating discussions with P. S\'{a}nchez-Friera, A.
Schindlmayr, R. Needs, U. von Barth, C.O. Ambladh, and L. Reining. P.G.-G.
thanks the Spanish Education Ministry for providing financial support
through a post-doctoral grant. 


\begin{table}[tbp]
\centering
\leavevmode
\begin{tabular}{lllllll}
\multicolumn{1}{c}{$r_{{\rm s}}$} & \multicolumn{1}{c}{1} & 
\multicolumn{1}{c}{2} & \multicolumn{1}{c}{4} & \multicolumn{1}{c}{5} & 
\multicolumn{1}{c}{10} & \multicolumn{1}{c}{20} \\ \hline
QMC & 0.5180 & 0.2742 & 0.1464 & 0.1197 & 0.0644 & 0.0344 \\ 
& 0.5127 & 0.2713 &  & 0.1201 &  & 0.0344 \\ 
$GW$ & 0.5160(2) & 0.2727(5) & 0.1450(5) & 0.1185(5) & 0.0620(9) & 0.032(1)
\\ 
&  & 0.2741 & 0.1465 &  &  &  \\ 
$GW_0$ & 0.5218(1) & 0.2736(1) & 0.1428(1) & 0.1158(1) & 0.0605(4) & 0.030(1)
\\ 
$G_0W_0$ & 0.5272(1) & 0.2821(1) & 0.1523(1) & 0.1247(1) & 0.0665(2) & 
0.0363(5) \\ 
RPA & 0.5370 & 0.2909 & 0.1613 & 0.1340 & 0.0764 & 0.0543 \vspace{2mm} \\ 
$-\varepsilon_{{\rm X}}$ & 0.4582 & 0.2291 & 0.1145 & 0.0916 & 0.0458 & 
0.0229
\end{tabular}
\caption{ Minus XC energies per particle (in Hartrees) for the
spin-unpolarized phase of the 3D homogeneous electron gas obtained through
several $GW$ schemes. The second row in the $GW$ entry corresponds to Ref.
[6]. Also shown are the RPA results, and the QMC values from Ref. [23]
(first row) and from Ref. [24] (second row). Parenthesis indicates the
numerical uncertainty in the last significant figure. For reference, the
exchange energy per particle, $\varepsilon_{{\rm X}}$, is included.}
\label{tab1}
\end{table}


\begin{figure}[tbp]
\caption{ XC energy per particle, $\varepsilon_{{\rm XC}}$, for the
spin-unpolarized 3D homogeneous electron gas. The essentially exact Monte
Carlo results (symbols) are compared with several $GW$ schemes (lines). The
excellent performance of the self-consistent $GW$ and (to a lesser extent)
the partially self-consistent $GW_0$, is evident. Note that the differences
between several Monte-Carlo results are less than the symbols size.}
\label{fig1}
\end{figure}

\begin{figure}[tbp]
\caption{ As Fig. \ref{fig1}, for the fully spin-polarized $\left( \chi =1
\right) $ phase of the 3D electron gas.}
\label{fig2}
\end{figure}

\begin{figure}[tbp]
\caption{ As in Fig. \ref{fig1}, for the spin-unpolarized 2D electron gas.}
\label{fig3}
\end{figure}


\vspace{-0.4cm}

\begin{references}
\bibitem{He65}  L. Hedin, Phys. Rev. {\bf 139}, A796 (1965); L. Hedin and S.
Lundqvist, in {\em Solid State Physics 23}, ed. by H. Ehrenreich, F. Seitz,
and D. Turnbull (Academic, New York, 1969).

\bibitem{Hy85}  e.g.: M.S. Hybertsen and S.G. Louie, Phys. Rev. Lett. {\bf 55%
}, 1418 (1985); R.W. Godby, M. Schl\"{u}ter, and L.J. Sham, {\em ibid,} {\bf %
56}, 2415 (1986); F. Aryasetiawan and O. Gunnarsson, Rep. Prog. Phys. {\bf 61%
}, 237 (1998).

\bibitem{GM58}  V. Galitskii and A. Migdal, Soviet Phys-JEPT {\bf 7}, 96
(1958).

\bibitem{Lu67}  B.I. Lundqvist, Phys. Kondens. Materie {\bf 6}, 206 (1967);
B. I. Lundqvist and V. Samathiyakatanit, {\em ibid,} {\bf 9}, 231 (1969).

\bibitem{Sh96}  E.L. Shirley, Phys. Rev. B {\bf 54}, 7758 (1996).

\bibitem{Ho99}  B. Holm, Phys. Rev. Lett. {\bf 83}, 788 (1999); B. Holm and
U. von Barth, Phys. Rev. B {\bf 57}, 2108 (1998).

\bibitem{KS65}  W. Kohn and L.J. Sham, Phys. Rev. {\bf 140}, A1133 (1965).

\bibitem{Do99}  W. Kohn, Y. Meir, and D. E. Makarov, Phys. Rev. Lett. {\bf 80%
}, 4153 (1998); J.F. Dobson and J. Wang, {\em ibid,} {\bf 82}, 2123 (1999).

\bibitem{GM97}  J. Grossman and L. Mitas, Phys. Rev. Lett. {\bf 79}, 4353
(1997).

\bibitem{Le99}  W. Leung, R.J. Needs, and G. Rajagopal, Phys. Rev. Lett. 
{\bf 83}, 2351 (1999).

\bibitem{Ki00}  Y. Kim, I. Lee, S. Nagaraja, J.P. Leburton, R.Q. Hood, and
R.M. Martin, Phys. Rev. B {\bf 61}, 5202 (2000); P. Garc\'{\i %
}a-Gonz\'{a}lez, {\em ibid }{\bf 62}, 2321 (2000).

\bibitem{Ne00}  W.M.C. Foulkes, L. Mitas, G. Rajagopal, and R.J. Needs, Rev.
Mod. Phys. (to be published).

\bibitem{Ro95}  N.H. Rojas, R.W. Godby, and R.J. Needs, Phys. Rev. Lett. 
{\bf 74}, 1827 (1995).

\bibitem{St99}  M.M. Rieger, L. Steinbeck, I.D. White, N.H. Rojas, and R.W.
Godby, Comput. Physics Commun. {\bf 117}, 211 (1999); L. Steinbeck, A.
Rubio, L. Reining, M. Torrent, I.D. White, and R.W. Godby, {\em ibid}, {\bf %
125}, 105 (2000).

\bibitem{Not1}  Note that $\Delta \mu =\mu -\mu _{0}$, where $\mu _{0}$ is
the chemical potential of the fictitious system under the potential $v_{0}$
alone.

\bibitem{Not2}  The factor 2 in Eq. \ref{4} comes out after a sum over the
spin variable in spin-unpolarized systems. For spin-polarized systems such
sum has to be done explicitly.

\bibitem{Sh97}  A. Schindlmayr, Phys. Rev. B {\bf 56}, 3528 (1997).

\bibitem{BK61}  G. Baym and L.P. Kadanoff, Phys. Rev. {\bf 124}, 287 (1961).

\bibitem{Ka62}  G. Baym, Phys. Rev. {\bf 127}, 1391 (1962).

\bibitem{BH96}  U. von Barth and B. Holm, Phys. Rev. B {\bf 54}, 8411 (1996).

\bibitem{SE98}  W.D. Sch\"{o}ne and A.G. Eguiluz, Phys. Rev. Lett. {\bf 81},
1662 (1998).

\bibitem{Hi97}  M. Hindgren, Ph.D. Thesis, University of Lund, 1997
(unpublished); U. von Barth (private communication).

\bibitem{CA80}  D.M. Ceperley and B.J. Adler, Phys. Rev. Lett. {\bf 45}, 566
(1980)

\bibitem{OB99}  G. Ortiz, M. Harris, and P. Ballone, Phys. Rev. Lett. {\bf 82%
}, 5317 (1999); G. Ortiz and P. Ballone, Phys. Rev. B {\bf 50}, 1391 (1994).

\bibitem{Bi82}  R.F. Bishop and K.H. L\"{u}hrmann, Phys. Rev. B {\bf 26},
5523 (1982).

\bibitem{TC90}  B. Tanatar and D. M. Ceperley, Phys. Rev. B {\bf 39}, 5005
(1989); Y. Kwon, D. M. Ceperley and R. Martin, {\em ibid,} {\bf 48}, 12037
(1993).

\bibitem{RG99}  M.M. Rieger and R.W. Godby, Phys. Rev. B {\bf 58}, 1343
(1998).
\end{references}
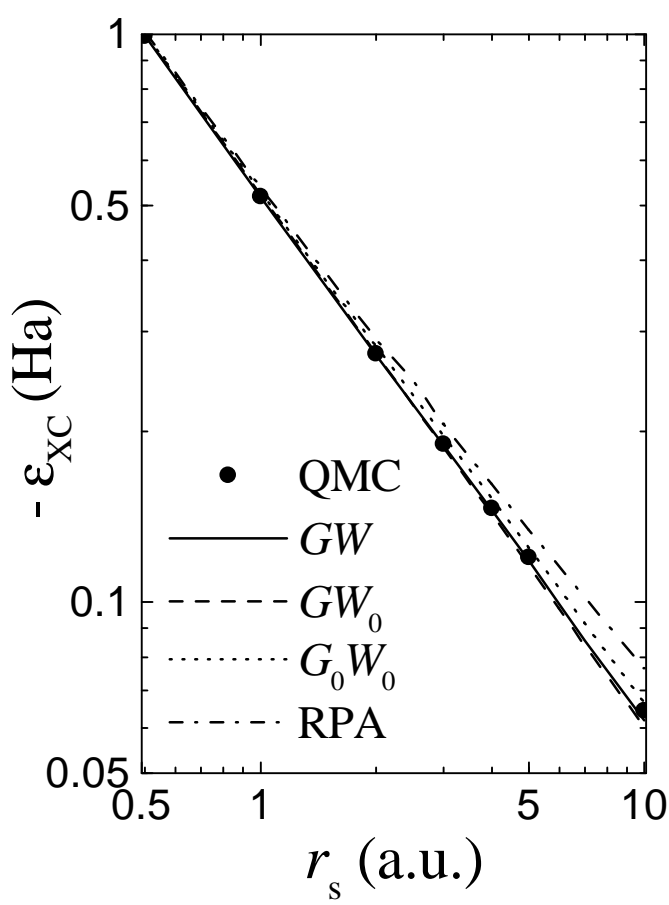
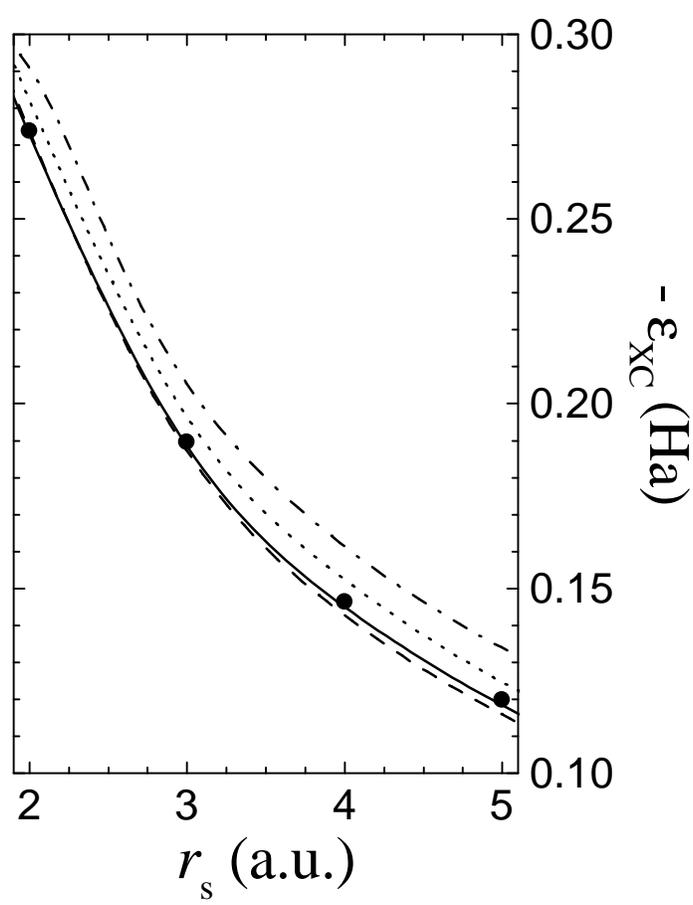
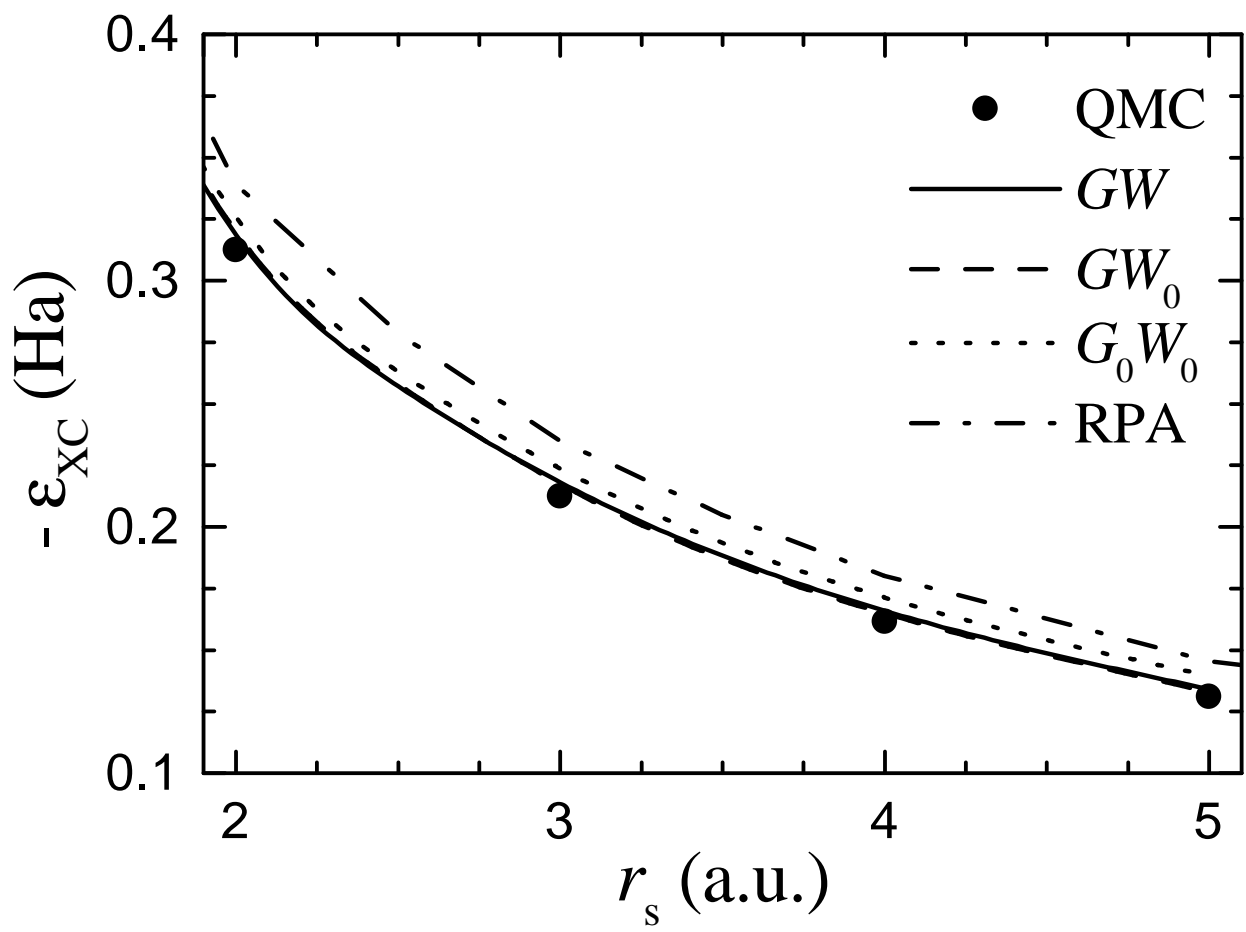
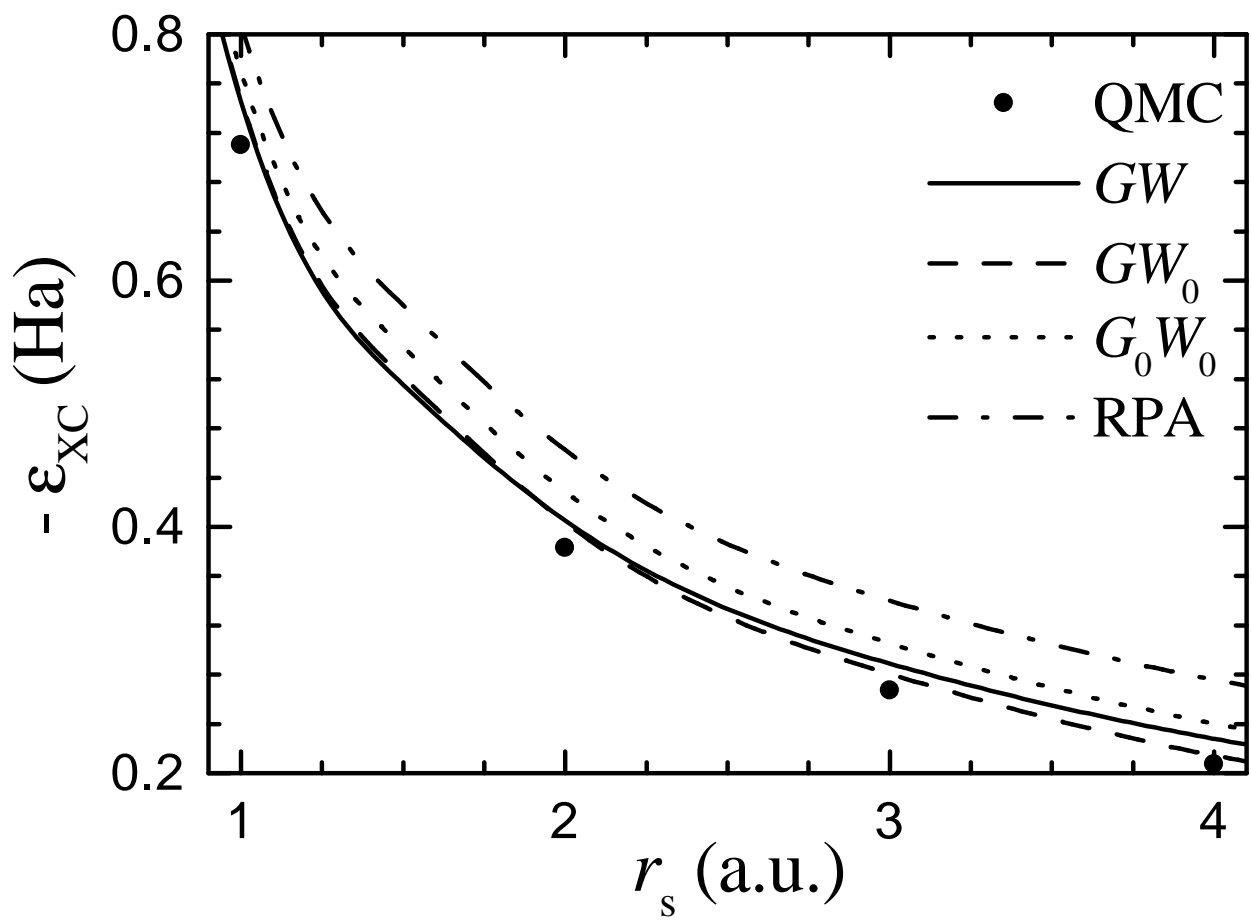
\end{document}